\documentclass[aps,prd,nopacs,floatfix,notitlepage,nofootinbib,twocolumn,a4paper,longbibliography]{revtex4-1}

\usepackage{amsfonts,amsmath,units,wasysym,epsfig,graphicx,verbatim,color,subfigure,graphicx,bm,mathrsfs,lipsum,hyperref,cleveref}
\usepackage{booktabs}
\usepackage[normalem]{ulem}  % \sout{old text} for strikeout
\usepackage{multirow}
\usepackage[utf8]{inputenc}

\begin{document}

\newcommand{\bhaskar}[1]{\textcolor{blue}{ \bf BB: #1}}
\newcommand{\pc}[1]{\textsf{\color{red}{ #1}}}

\newcommand{\HU}{{Hamburger Sternwarte, Gojenbergsweg 112, D-21029 Hamburg, Germany}}

\newcommand{\USAL}{Departamento de F\'isica Fundamental and IUFFyM, Universidad de Salamanca, Plaza de la Merced S/N, E-37008 Salamanca, Spain}
\newcommand{\Uliege}{Space Sciences, Technologies and Astrophysics Research (STAR) Institute, Universit\'e de Li\`ege, B\^at. B5a, 4000 Li\`ege, Belgium}

\title{Systematics from NICER Pulse Profiles Drive Uncertainty in Multi-Messenger Inference of the Neutron Star Equation of State}

\author{Bhaskar Biswas$^{\rm 1}$, Prasanta Char$^{\rm 2, 3}$}
\affiliation{$^{\rm 1}$\HU, $^{\rm 2}$\USAL, $^{\rm 3}$\Uliege}

\begin{abstract}
We present new constraints on the neutron star equation of state (EOS) and mass distribution using a unified Bayesian inference framework that incorporates latest NICER measurements, including PSR J0614$-$3329, alongside gravitational wave data, radio pulsar masses, and nuclear theory. By systematically comparing four inference scenarios—varying in the inclusion of PSR J0614$-$3329 and in the pulse profile model used for PSR J0030+0451—we quantify the impact of observational and modeling choices on dense matter inference. We find that pulse profile systematics dominate EOS uncertainties: the choice of hot spot geometry for PSR J0030+0451 leads to significant shifts in the inferred stiffness of the EOS and maximum neutron star mass. In contrast, PSR J0614$-$3329 mildly softens the EOS at low densities, reducing the radius at \(1.4\,M_\odot\) by \(\sim 100\)~m. A Bayesian model comparison yields a Bayes factor of $\log_{10} \mathrm{BF} \approx 1.58$ in favor of the ST+PDT model over PDT-U, providing strong evidence that multi-messenger EOS inference can statistically discriminate between competing NICER pulse profile models.  These results highlight the critical role of NICER systematics in dense matter inference and the power of joint analyses in breaking modeling degeneracies.
\end{abstract}

\maketitle

\section{Introduction}

Understanding the equation of state (EOS) of dense matter is a central problem in modern astrophysics and nuclear physics. The EOS encodes the relationship between pressure and energy density in the interiors of neutron stars (NSs), where matter reaches densities several times that of atomic nuclei~\cite{Lattimer_2016,Oertel_2017,Baym_2018}. Because such conditions cannot be reproduced in terrestrial experiments, neutron stars serve as unique astrophysical laboratories for studying the behavior of strongly interacting matter under extreme conditions.

Over the past decade, rapid advances in both theory and observation have transformed our ability to constrain the EOS. On the observational side, precise measurements of neutron star masses and radii have been obtained from X-ray observations~\citep{Riley:2019yda,Miller:2019cac,Riley:2021pdl,Miller:2021qha,Choudhury:2024xbk} (particularly from the Neutron Star Interior Composition Explorer, NICER~\citep{2016SPIE.9905E..1HG_2}), and gravitational wave (GW) signals from binary neutron star (BNS) mergers~\citep{TheLIGOScientific:2017qsa,Abbott:2018wiz,Abbott:2018exr,Abbott:2020uma} observed by LIGO/Virgo/KAGRA (LVK) collaboration ~\citep{advanced-ligo,advanced-virgo}. On the theoretical side, nuclear many-body calculations based on chiral effective field theory (\(\chi\)EFT)~\cite{Epelbaum:2008ga,Machleidt:2011zz,Hammer:2012id,Hebeler:2020ocj,Drischler:2021kxf} and perturbative Quantum Chromodynamics (pQCD)~\cite{Komoltsev:2021jzg} have provided reliable predictions at low and high densities, respectively, while experimental efforts such as PREX-II \cite{PREX:2021umo} and CREX \cite{CREX:2022kgg} have yielded complementary constraints on nuclear symmetry energy parameters.

In the preceding study, \textcite{Biswas:2024hja} focused on constraining the equation of state (EOS) for neutron stars (NS) using a multifaceted approach that integrates theoretical models, observational data, and experimental results. This comprehensive effort aimed to improve our understanding of the internal structure and composition of neutron stars, particularly under the extreme conditions found within them.

The authors employed a hybrid EOS model~\cite{Biswas:2020puz,Biswas:2020xna,Biswas:2021yge} that combines a parabolic expansion-based nuclear empirical parameterization~\cite{PhysRevC.44.1892,Haensel:2007yy,Baldo:2016jhp, Margueron:2017eqc} around nuclear saturation density along with a three-segment piecewise polytrope~\cite{Read:2008iy} model at higher densities. This approach leverages the strengths of both models to provide a robust description of the EOS across a wide range of densities. Key data sources incorporated into the analysis included theoretical models, such as, \(\chi\)EFT for low to moderate densities and pQCD for high densities. Experimental results from PREX-II and CREX provided measurements related to the neutron skin thickness of heavy nuclei. Observational data comprised a total of 130 neutron star mass measurements up to April 2023~\cite{Fan:2023spm}, simultaneous mass and radius measurements of pulsars PSR J0030+0451 \cite{Riley:2019yda,Miller:2019cac}, PSR J0740+6620 \cite{Riley:2021pdl, Miller:2021qha}, and PSR J0437+4715 \cite{Choudhury:2024xbk} from the NICER, and constraints from tidal deformabilities inferred from binary neutron star mergers GW170817 \cite{Abbott:2018wiz,Abbott:2018exr} and GW190425 \cite{Abbott:2020uma}, observed by the LIGO/Virgo/Kagra collaboration. A hierarchical Bayesian framework was employed to integrate these diverse constraints, allowing for the simultaneous inference of the EOS and the NS mass distribution.

The study found that incorporating data from \(\chi\)EFT significantly tightened the constraints on the EOS near or below nuclear saturation density. In contrast, constraints derived from pQCD and nuclear experiments (PREX-II and CREX) had minimal impact on the EOS. Key inferred EOS parameters included the slope (L) of the symmetry energy, determined to be $54^{+10}_{-10}$ MeV (with a 90\% credibility interval (CI)), and the curvature (\(K_{\text{sym}}\)) of the symmetry energy, determined to be $-158^{+73}_{-63}$ MeV (with a 90\% CI). Inferred neutron star properties included a radius of $12.34_{- 0.53}^{+0.43}$ km and a tidal deformability of $436_{-117}^{+109}$ for a 1.4 solar mass NS, and a maximum mass of $2.22_{-0.19}^{+0.21} M_{\odot}$  for a non-rotating NS (with a 90\% CI).

Despite these advances, a critical open question remains: \textit{\textbf{How sensitive are EOS inferences to the modeling assumptions underlying specific observational inputs—particularly NICER X-ray pulse profile analyses, which depend on assumed hotspot geometries and emission models?}} In recent reanalyses of NICER data, multiple viable pulse profile models have emerged, often yielding substantially different radius and mass estimates for the same neutron star (e.g., PSR J0030+0451)~\cite{Vinciguerra:2023qxq}. This raises important concerns about model dependence and degeneracy in EOS inference frameworks.

In this work, we assess whether a multi-messenger, hierarchical Bayesian EOS analysis—integrating constraints from NICER, gravitational waves, radio pulsar masses, nuclear theory, and laboratory experiments—\textit{\textbf{can statistically distinguish between competing pulse profile models.}} Specifically, we perform joint EOS and population inference under two alternative pulse profile assumptions for PSR J0030+0451: the standard two-spot (\texttt{ST+PDT}) configuration and a more complex, unconstrained multi-hotspot geometry (\texttt{PDT-U}). \textcite{Vinciguerra:2023qxq} has reported that both \texttt{ST+PDT} and \texttt{PDT-U} models fit both the NICER-only and the combined NICER and XMM-Newton data, reasonably well. The \texttt{ST+PDT} model estimates the mass and the radius of PSR J0030+0451 to $1.40^{+0.13}_{-0.12} M_\odot$ and  $11.71^{+0.88}_{-0.83}$ km, respectively. However, for the \texttt{PDT-U} model, the corresponding solutions are $1.70^{+0.18}_{-0.19} M_\odot$ and $14.44^{+0.88}_{-1.05}$ km, respectively. Hence, we will use the publicly available posterior samples for both these solutions \cite{vinciguerra_2023_8239000} to investigate their impact on the EOS inference exercise. In our previous work~\cite{Biswas:2024hja}, we examined both pulse profile scenarios and noted a significant difference in the inferred mass-radius posteriors. However, a detailed investigation was not pursued to quantify the corresponding differences in the inferred EOS parameters, macroscopic neutron star properties, or to compare the Bayesian evidence formally between these competing models. The primary aim of this paper is to conduct a comprehensive analysis addressing these aspects in detail.

We also incorporate the recent NICER mass–radius posterior for PSR J0614$-$3329~\cite{Mauviard:2025dmd}, a recycled millisecond pulsar whose relatively small inferred radius at moderate mass provides a potentially powerful constraint on the EOS. By including this new measurement, we test its impact on the pressure–density relation, radius predictions, and the consistency of existing EOS constraints.

Our analysis shows that the inferred EOS, macroscopic neutron star properties (e.g., radius and mass of 1.4 $M_{\odot}$ NS, maximum mass), and pressure--density relations are highly sensitive to the adopted pulse profile model for PSR J0030+0451. Moreover, Bayesian model comparison yields a Bayes factor of approximately 44 ($\Delta \log Z \approx 3.63$), strongly favoring the \texttt{ST+PDT} model over \texttt{PDT-U} when all data are considered in a unified framework. This demonstrates that multi-messenger EOS inference can effectively discriminate between competing geometric interpretations of X-ray pulse profiles, providing an important consistency check on NICER modeling assumptions.

\section{Inference Methodology}
\label{section: Methodology}

Following our previous work~\cite{Biswas:2024hja}, we employ a hierarchical Bayesian framework to jointly constrain the neutron star EOS and mass distribution. This approach integrates diverse datasets spanning theory, experiment, and observation. The posterior distribution of the model parameters is obtained via Bayes' theorem:

\[
P(\boldsymbol{\theta} \mid D) = \frac{P(D \mid \boldsymbol{\theta})\, P(\boldsymbol{\theta})}{P(D)},
\]

where $\boldsymbol{\theta}$ represents the set of model parameters—including EOS and mass distribution parameters—and $D$ denotes the combined dataset. The likelihood $P(D \mid \boldsymbol{\theta})$ encodes the agreement between model predictions and data, while $P(\boldsymbol{\theta})$ represents prior knowledge or theoretical constraints. The evidence $P(D)$ acts as a normalization constant and is not required for parameter estimation.

The EOS is parametrized using a hybrid approach combining a low-density empirical parameterization with a three-segment piecewise polytrope at higher densities. The NS mass distribution is modeled as a two-component Gaussian mixture to capture the observed diversity in mass measurements.

The full set of model parameters \( \boldsymbol{\theta} \) inferred in this work includes both EOS and neutron star mass distribution parameters.

\paragraph*{~\textbf{EOS Parameters.}}
The EOS is constructed using a hybrid approach that connects a low-density segment based on the SLy EOS~\cite{Douchin:2001sv} to an empirical nuclear parameterization, which is in turn matched to a high-density three-segment piecewise polytrope. Continuity in pressure is enforced at each transition. The parameters are:

\begin{itemize}
    \item \( L \): slope of the symmetry energy at saturation density,
    \item \( K_{\text{sym}} \): curvature of the symmetry energy,
    \item \( n_1 \): transition density between the empirical segment and the piecewise polytrope,
    \item \( n_2, n_3 \): transition densities between the polytropic segments,
    \item \( \Gamma_1, \Gamma_2, \Gamma_3 \): adiabatic indices (polytropic exponents) defining the pressure–density relation in each segment.
\end{itemize}

The constructed EOS satisfies fundamental physical constraints, including:
\begin{itemize}
    \item \textit{Causality}, i.e., the speed of sound does not exceed the speed of light;
    \item \textit{Thermodynamic stability}, enforced through monotonic increase of pressure with density.
\end{itemize}

\paragraph*{~\textbf{Mass Distribution Parameters.}}
The neutron star mass distribution is modeled as a mixture of two Gaussian components, representing different sub-populations. The corresponding parameters are:
\begin{itemize}
    \item \( \mu_1, \mu_2 \): the means of the two Gaussian components,
    \item \( \sigma_1, \sigma_2 \): their standard deviations,
    \item \( w \): the mixture fraction associated with the first component (with \( 1 - w \) assigned to the second).
\end{itemize}

\paragraph*{~\textbf{Prior Ranges.}}
We adopt the same prior distributions as in our previous work~\cite{Biswas:2024hja}, where all EOS and mass distribution parameters were assigned uniform priors within physically motivated ranges. The full list of parameters and their respective prior bounds is summarized in Table 1 of~\cite{Biswas:2024hja}.

\paragraph*{~\textbf{Likelihood .}} The likelihood function is constructed by combining:
\begin{itemize}
    \item NICER’s simultaneous mass–radius inferences for PSRs J0030+0451~\cite{Vinciguerra:2023qxq, vinciguerra_2023_8239000}, J0740+6620~\cite{Salmi:2024aum, salmi_2024_10519473}, and J0437--4715~\cite{Choudhury:2024xbk, choudhury_2024_13766753};
    \item Tidal deformability posteriors from GW170817~\cite{Abbott:2018wiz} and GW190425~\cite{LIGOScientific:2020aai};
    \item Theoretical priors from chiral effective field theory (\(\chi\)EFT)~\cite{BUQEYEgithub,Drischler:2020hwi} and perturbative QCD (pQCD)\cite{Gorda:2021znl,Gorda:2022jvk,Komoltsev:2021jzg}, relevant at low and high densities, respectively;
    \item Neutron skin thickness measurements from PREX-II~\cite{PREX:2021umo} and CREX~\cite{CREX:2022kgg}, connected to symmetry energy parameters;
    \item There are three classes of neutron star mass measurements~\cite{Alsing:2017bbc,Fan:2023spm}. The first includes systems with individual mass measurements from Shapiro delay or relativistic orbital parameters, modeled using symmetric or asymmetric Gaussian likelihoods depending on the measurement uncertainties. The second and third classes consist of binary systems with measurements of either the total mass or the mass ratio (with associated uncertainties), combined with the mass function to infer neutron star masses by integrating over the inclination angle.
\end{itemize}

Here, we do not include the data from PSR J1231$-$1411 to provide a cleaner comparison, unlike some of earlier works that included it \cite{Li:2024pts,Li:2025oxi}. Moreover, the mass-radius measurements of PSR J1231$-$1411, being a challenging source to model, has issues with convergence of the Bayesian analysis and dependence on radius priors \cite{Salmi:2024bss}. 

Posterior samples are drawn using using nested sampling algorithm implemented in~{\tt PyMultiNest}~\citep{Buchner:2014nha}. Their constraints and correlations are presented in the following sections.

\begin{figure*}[ht!]
    \centering
    \includegraphics[width=\textwidth]{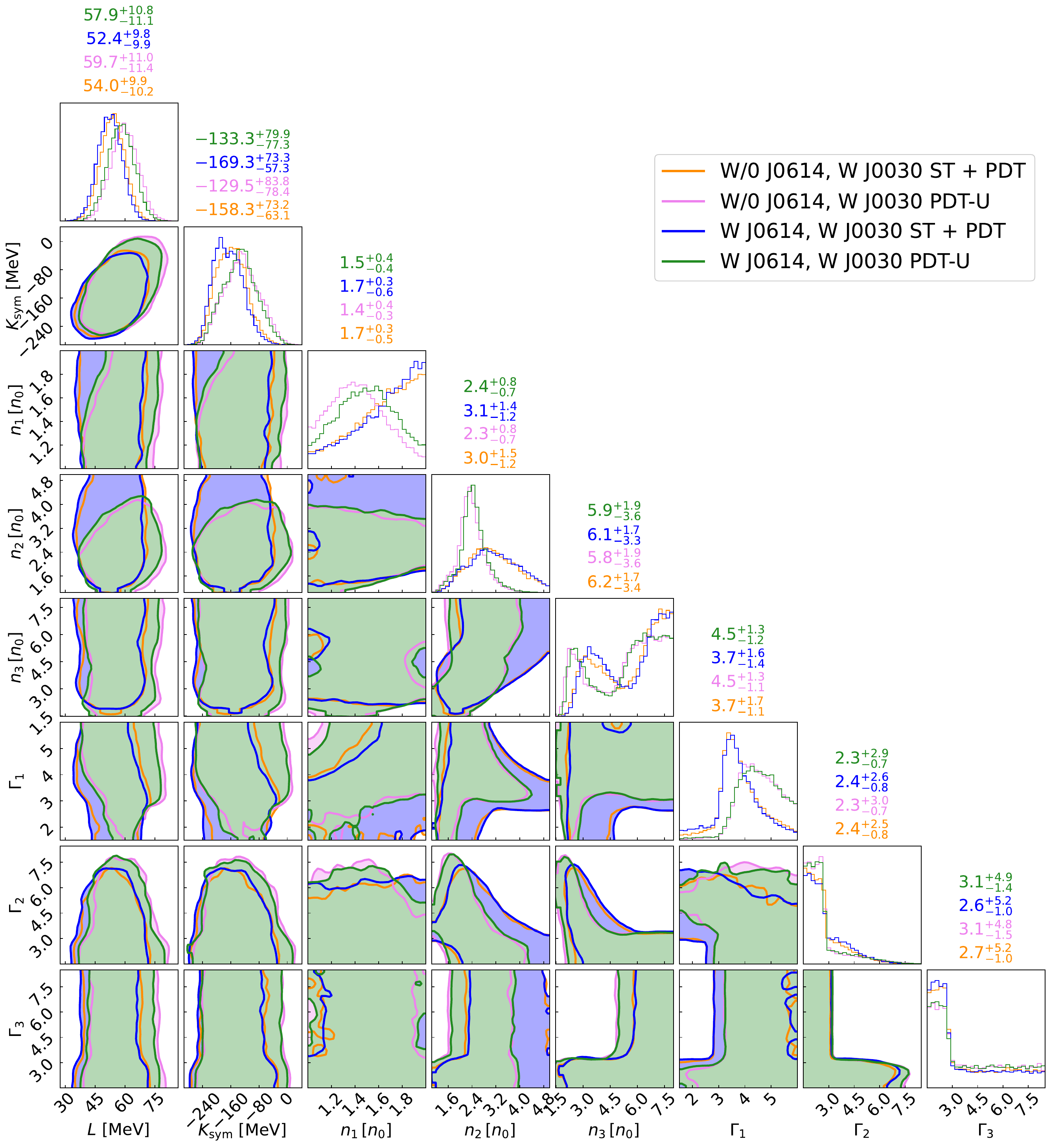}
    \caption{Posterior distributions of the eight EOS parameters across the four inference scenarios:
\texttt{W/O J0614, W J0030 ST + PDT} (blue),
\texttt{W J0614, W J0030 ST + PDT} (orange),
\texttt{W/O J0614, W J0030 PDT-U} (green),
and \texttt{W J0614, W J0030 PDT-U} (red).
Shown are the empirical nuclear parameters—slope \(L\) and curvature \(K_{\text{sym}}\) of the symmetry energy;
the three transition densities \(n_1\), \(n_2\), \(n_3\) (in units of nuclear saturation density \(n_0\));
and the three polytropic indices \(\Gamma_1\), \(\Gamma_2\), and \(\Gamma_3\) that describe the high-density EOS.
Both the inclusion of PSR J0614$-$3329 and the choice of pulse profile model for PSR J0030+0451 impact these parameters,
particularly the high-density indices and transition densities. In the marginalized one-dimensional plots, the median and 90\% CIs are shown for each constraint type, with each constraint represented in its respective color.}
    \label{fig:eos_params}
\end{figure*}

\begin{figure}[ht!]
    \centering

    \includegraphics[width=0.45\textwidth]{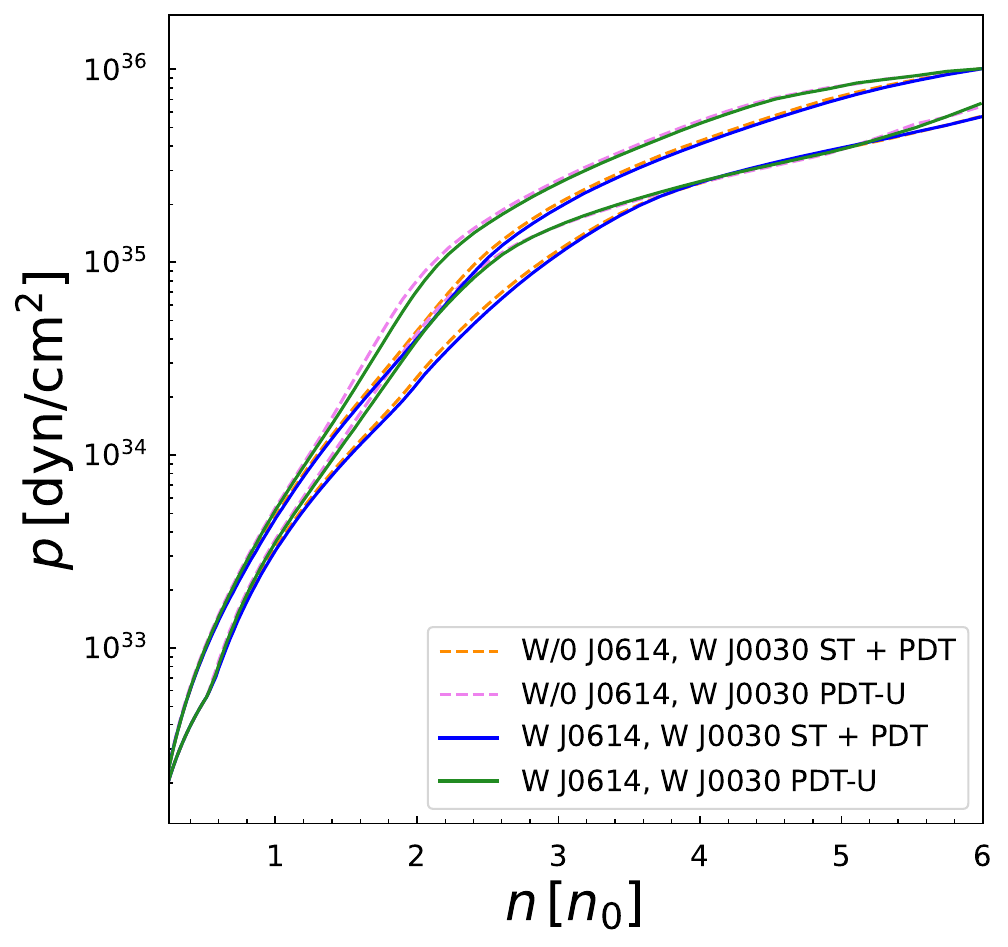} 
\caption{90 \% CI of the marginalized posterior distribution of the pressure in NS interior as a function of baryon density is shown for each scenario considered in this study, as indicated in the legend of the plot.}
    \label{fig:eos-post}
\end{figure}

\begin{figure}[ht!]
    \centering
    \includegraphics[width=0.45\textwidth]{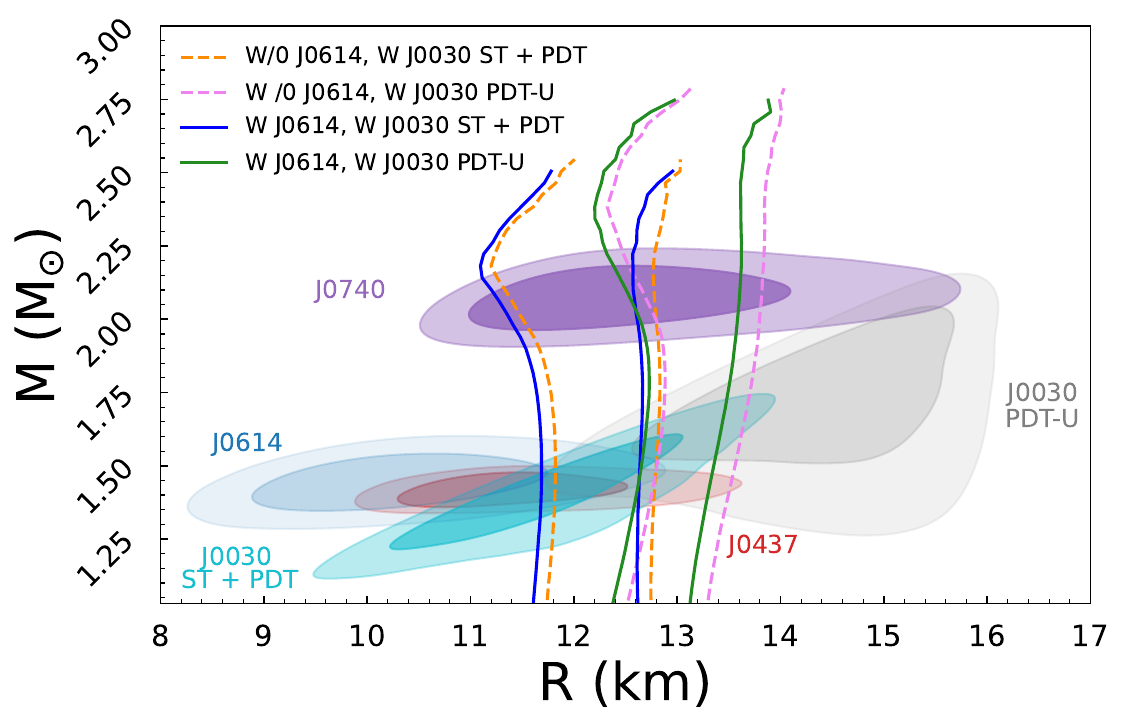}
  
\caption{90 \% CI of the marginalized posterior distribution of mass vs radius is shown for each scenario considered in this study, as indicated in the legend of the plot.}
    \label{fig:mr-post}
\end{figure}

\begin{figure*}[ht!]
    \centering
    \includegraphics[width=\textwidth]{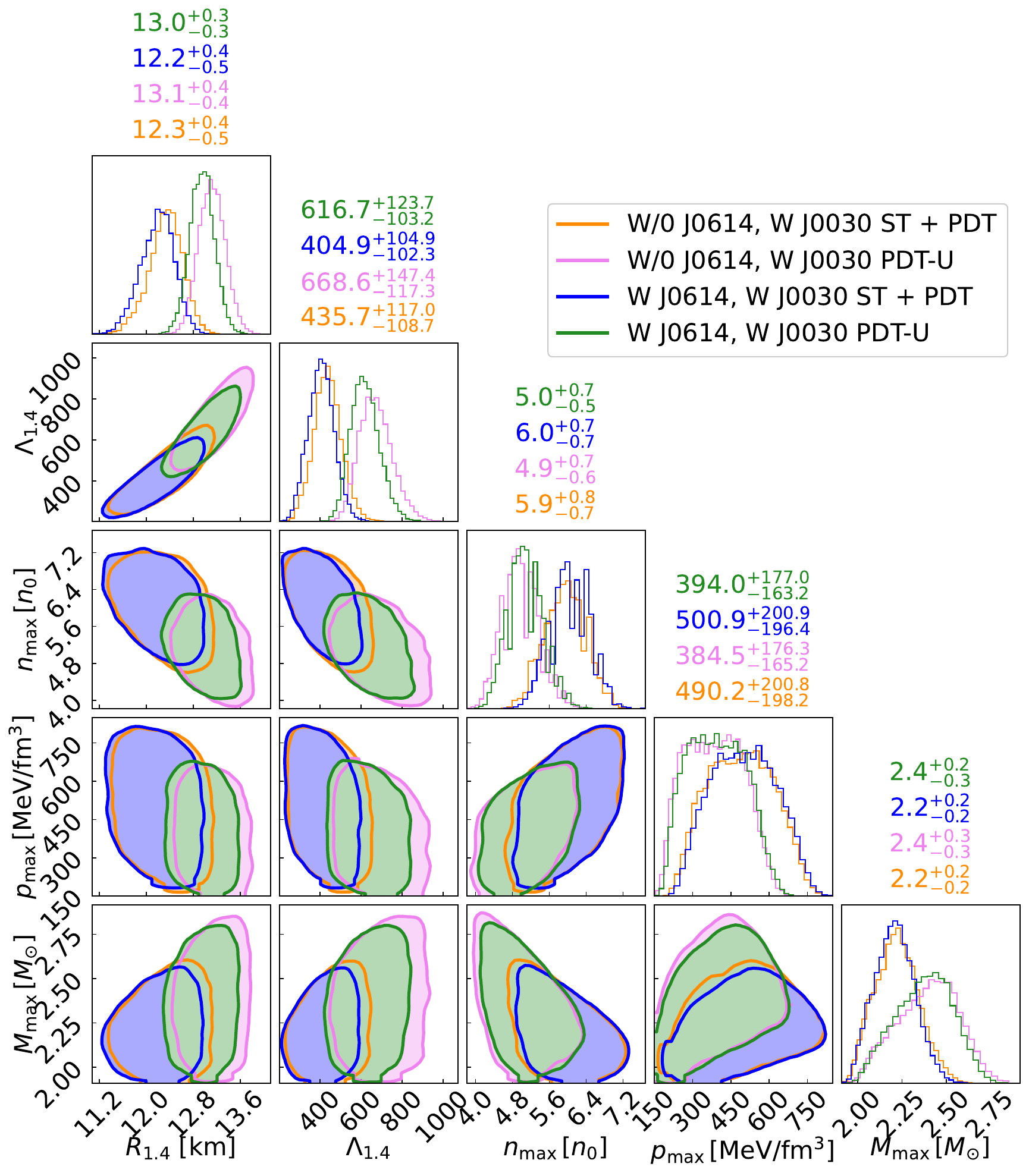}
\caption{Correlations between radius ($R_{1.4}$) and tidal deformability ($\Lambda_{1.4}$) of a $1.4M_{\odot}$ NS, maximum density ($n_{\rm max}$), maximum pressure ($p_{\rm max}$), and maximum mass ($M_{\rm max}$) are shown for each scenario considered in this study, as indicated in the legend of the plot. In the marginalized one-dimensional plots, the median and 90\% CIs are shown for each constraint type, with each constraint represented in its respective color.}
    \label{fig:macro_params}
\end{figure*}

\begin{figure*}[ht!]
    \centering
    \includegraphics[width=\textwidth]{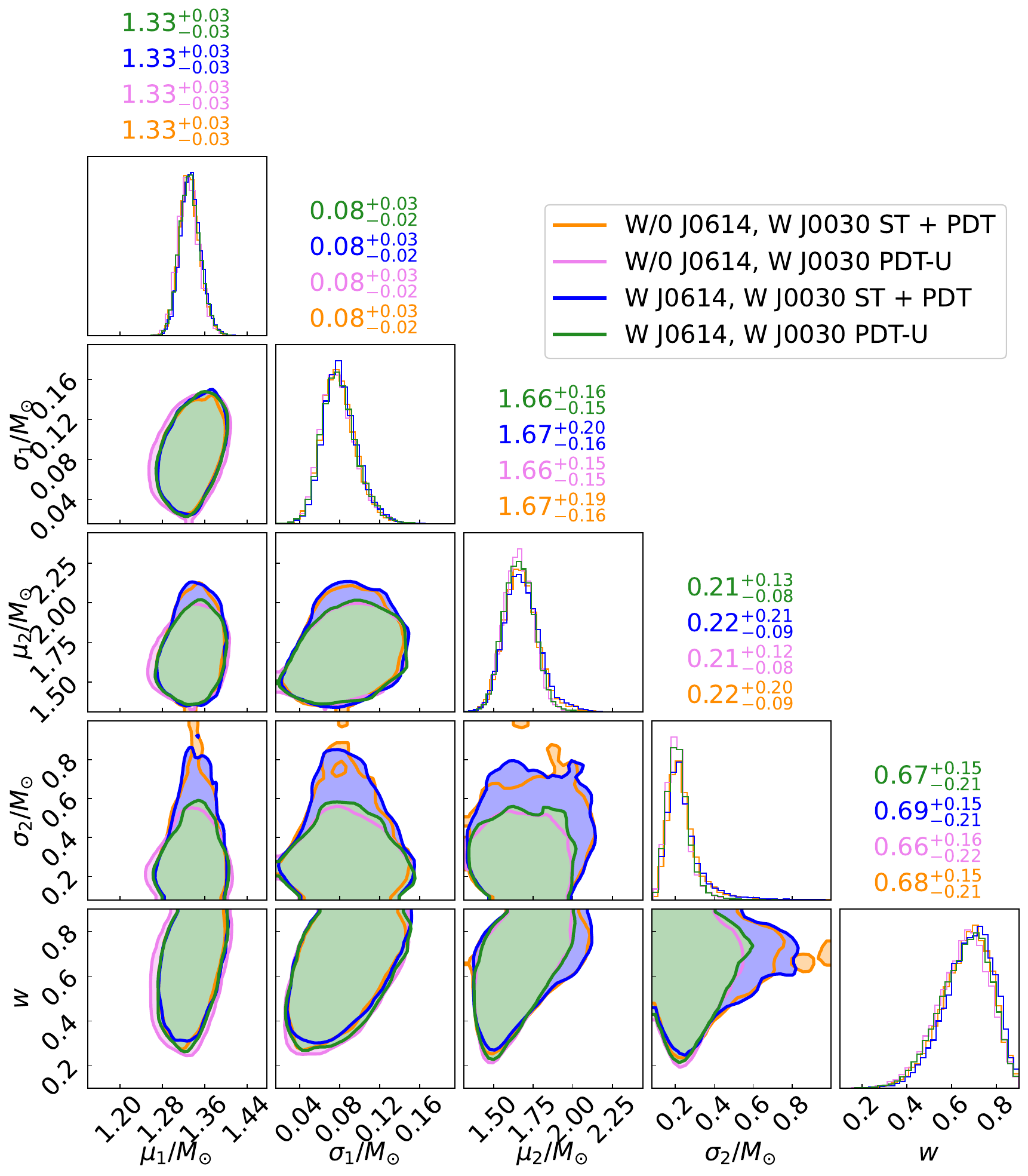}
\caption{Posterior distribution of mass model parameters are shown for each scenario considered in this study, as indicated in the legend of the plot. In the marginalized one-dimensional plots, the median and 90\% CIs are shown for each constraint type, with each constraint represented in its respective color.}
    \label{fig:pop_params}
\end{figure*}

\section{Results}
\label{section: results}

In this section, we present a comparative analysis of four inference scenarios designed to evaluate how recent observational inputs—specifically the NICER mass–radius measurement of PSR J0614$-$3329 and the re-analysis of PSR J0030+0451—affect constraints on the neutron star equation of state (EOS) and mass distribution. The four scenarios differ in whether they include the PSR J0614$-$3329 data and in which pulse profile model is adopted for PSR J0030+0451.

For clarity, figures in this section use the abbreviations \textbf{``W''} for \textit{with} and \textbf{``W/O''} for \textit{without} a given data component. The four scenarios are as follows:

\begin{enumerate}
    \item \textbf{Baseline Scenario} (\texttt{W/O J0614, W J0030 ST + PDT}): Includes mass–tidal deformability ($M$–$\Lambda$) posteriors from GW170817 and GW190425, NICER mass–radius inferences for PSRs J0030+0451 (using the \texttt{ST+PDT} model from \citet{Vinciguerra:2023qxq}), J0740+6620, and J0437+4715, along with 70 well-vetted radio pulsar mass measurements. Theoretical priors from chiral effective field theory ($\chi$EFT) and perturbative QCD (pQCD), as well as neutron skin thickness constraints from PREX-II and CREX, are also included.

    \item \textbf{Baseline + PSR J0614$-$3329} (\texttt{W J0614, W J0030 ST + PDT}): Extends the Baseline by adding the NICER mass–radius posterior of PSR J0614$-$3329.

    \item \textbf{Alternative PSR J0030+0451 Model} (\texttt{W/O J0614, W J0030 PDT-U}): Uses the same data as the Baseline but replaces the \texttt{ST+PDT} model for PSR J0030+0451 with the \texttt{PDT-U} model from \citet{Vinciguerra:2023qxq}, which assumes more complex hot spot geometries. This model results in a significantly larger inferred radius ($R = 14.44^{+0.88}_{-1.05} \, \mathrm{km}$) and higher mass ($M = 1.70^{+0.18}_{-0.19} \, M_{\odot}$).

    \item \textbf{PDT-U + PSR J0614$-$3329} (\texttt{W J0614, W J0030 PDT-U}): Combines the \texttt{PDT-U} model for PSR J0030+0451 with the NICER observation of PSR J0614$-$3329.
\end{enumerate}

Figure~\ref{fig:eos_params} shows the marginalized posterior distributions of the eight EOS parameters used in our model:
the slope (\(L\)) and curvature (\(K_{\text{sym}}\)) of the symmetry energy,
the transition densities \(n_1\), \(n_2\), \(n_3\), and the polytropic indices \(\Gamma_1\), \(\Gamma_2\), \(\Gamma_3\)
governing the three piecewise segments above nuclear saturation density. The low-density empirical parameters \(L\) and \(K_{\text{sym}}\) are primarily influenced by the $M$–$\Lambda$ posteriors from GW170817,  mass–radius posteriors of PSRs J0030+0451 and J0437+4715, and to a lesser extent by the theoretical priors from \(\chi\)EFT. The use of the \texttt{PDT-U} model for PSR J0030+0451 consistently shifts the \(L\) and \(K_{\text{sym}}\) posterior to higher values, reflecting the requirement of higher pressure to support larger radii. In contrast, the inclusion of PSR J0614$-$3329 has only a mild effect on \(L\) and \(K_{\text{sym}}\), but it shifts the posteriors slightly towards the smaller values.

The transition densities \(n_1\), \(n_2\), and \(n_3\), as well as the polytropic index \(\Gamma_1\), are noticeably impacted by the observational inputs. The inclusion of PSR J0614$-$3329 primarily affects \(n_1\), the density at which the empirical EOS transitions into the first polytropic segment. This is expected, as J0614’s relatively small radius ($R = 10.29^{+1.01}_{-0.86}\,\mathrm{km}$) at a modest mass ($M = 1.44^{+0.06}_{-0.07}\,M_\odot$) provides important constraints on the pressure at low to intermediate densities, tightening the posterior on the onset of the polytropic regime. In contrast, the \texttt{PDT-U} model for PSR J0030+0451 has a broader impact on the EOS structure, shifting all three transition densities (\(n_1\), \(n_2\), and \(n_3\)) and modifying the slope of the first polytropic segment through \(\Gamma_1\). This is consistent with the need for a stiffer EOS to support the larger radius inferred under this model, especially at masses around $1.7\,M_\odot$. For completeness, we include in Figure~\ref{fig:eos-post} the reconstructed pressure–density relations corresponding to the four inference scenarios, derived from the EOS parameter posteriors shown in Figure~\ref{fig:eos_params}. We have shown the $90\%$ CI of the EOS for all cases. We can see from Figure \ref{fig:eos-post} that the EOS posterior for the Baseline case around $2n_0$ is softer than the Alternative case without PSR J0614$-$3329. This behavior is due to the requirement of a larger radius for an intermediate mass NS for the  \texttt{PDT-U} solution. With the addition of J0614$-$3329, we see a general softening, however small, at all densities for both Baseline and Alternative cases.

Figure~\ref{fig:mr-post} shows the inferred mass–radius relations for neutron stars under all four inference scenarios, corresponding to the EOS posteriors shown in Figure \ref{fig:eos-post}. The $90\%$ credible intervals are represented by the region enclosed between solid and dashed lines, and the NICER posteriors for individual sources are plotted for reference. The mass-radius posteriors closely follow the behavior of the EOS posteriors. The most prominent feature is the contrast between scenarios using the \texttt{ST+PDT} and \texttt{PDT-U} models for PSR J0030+0451. The \texttt{PDT-U} model (solid green and dashed violet curves), which infers a larger radius for a \(1.7\,M_\odot\) star, leads to a visibly stiffer \(M\)–\(R\) relation. Radii at \(1.4–1.7\,M_\odot\) are typically larger by \(0.5–1.0\,\mathrm{km}\) compared to the \texttt{ST+PDT} counterparts.
The inclusion of PSR J0614$-$3329 (solid blue and solid green curves) reduces the uncertainty at the high-mass end (\(M \gtrsim 1.8\,M_\odot\)) and slightly lowers ($\sim 100$ m) the radius at moderate masses. This is consistent with its relatively small inferred radius at \(1.44\,M_\odot\), which pulls the allowed EOS slightly softer in that regime. 

Figure~\ref{fig:macro_params} presents the posterior distributions of key macroscopic neutron star observables inferred in each scenario: the radius \(R_{1.4}\) and tidal deformability \(\Lambda_{1.4}\) of a 1.4\,\(M_\odot\) neutron star, the maximum non-rotating mass \(M_{\text{max}}\), and the central pressure \(p_{\text{max}}\) and central density \(n_{\text{max}}\) associated with each configuration. The values of \(R_{1.4}\) and \(\Lambda_{1.4}\) are highly sensitive to the pulse profile model adopted for PSR J0030+0451. Scenarios using the \texttt{PDT-U} model yield systematically larger radii (\( \sim 13.0\text{--}13.1\,\mathrm{km} \)) and higher deformabilities (\( \Lambda_{1.4} \sim 669 \)) compared to those using the \texttt{ST+PDT} model (\( \sim 12.2\text{--}12.3\,\mathrm{km} \), \( \Lambda_{1.4} \sim 405 \)), reflecting the stiffer EOS required to support the larger inferred radius under \texttt{PDT-U}.

The inclusion of PSR J0614$-$3329, which has a relatively small radius at a mass of \(1.44\,M_\odot\), softens the EOS at low to intermediate densities. As a result, the median value of \(R_{1.4}\) is reduced by approximately 100 meters across both \texttt{ST+PDT} and \texttt{PDT-U} configurations. The inferred radius of PSR J0614$-$3329 is consistent with those of PSR J0030+0451 (\texttt{ST+PDT}), PSR J0437$-$4715, and the binary neutron star merger GW170817.

In contrast, the increase in \(M_{\text{max}}\), as well as in \(p_{\text{max}}\) and \(n_{\text{max}}\), is primarily driven by the use of the \texttt{PDT-U} model. The stiffer EOS inferred in this case supports a higher maximum mass (from \(\sim 2.2\) to \(\sim 2.4\,M_\odot\)) and leads to correspondingly smaller central densities and pressures. The addition of PSR J0614$-$3329 has only a secondary effect on these quantities, primarily sharpening the posterior distributions without significantly shifting their peaks.

Figure~\ref{fig:pop_params} displays the joint and marginalized posterior distributions for the parameters of the neutron star mass distribution, modeled as a two-component Gaussian mixture. The parameters include the means (\(\mu_1\), \(\mu_2\)), standard deviations (\(\sigma_1\), \(\sigma_2\)), and mixture weight (\(w\)) of the lower-mass component. We find that the mass distribution parameters remain remarkably consistent across all four inference scenarios. The primary component, centered around \(\mu_1 \approx 1.33\,M_\odot\), is tightly constrained with a narrow spread of \(\sigma_1 \sim 0.08\,M_\odot\). The secondary component is broader, centered around \(\mu_2 \approx 1.66\,M_\odot\), with \(\sigma_2 \sim 0.21\,M_\odot\), and constitutes a smaller fraction of the population. Neither the inclusion of PSR J0614$-$3329 nor the choice of pulse profile model for PSR J0030+0451 has any noticeable effect on these parameters. This stability indicates that the mass distribution is primarily informed by the 70 radio pulsar mass measurements included in all scenarios, and is largely independent of the  NICER and gravitational wave data.

\subsection{Model Comparison via Bayesian Evidence}

To assess the relative statistical support for different pulse profile models of PSR J0030+0451, we compute the Bayesian model evidence \(\log Z\) for each configuration. We obtain:
\begin{align*}
\log Z_{\text{ST+PDT}} &= -11.819 \pm 0.015 \\
\log Z_{\text{PDT-U}}  &= -15.449 \pm 0.011
\end{align*}

The resulting difference, \(\Delta \log Z = 3.63 \pm 0.019\), corresponds to a Bayes factor of approximately \( \log_{10} \text{BF} \approx 1.58 \), which constitutes \emph{strong} evidence in favor of the \texttt{ST+PDT} model, following the interpretation of Kass and Raftery~\citep{Kass:1995loi}. Although not formally decisive by their criteria, this level of support is still strong. Models with \( \log_{10} Z \leq -2 \) compared to the best-fitting model can be considered decisively ruled out.

This result implies that, despite the improved pulse profile fits achieved by the more complex \texttt{PDT-U} model in standalone NICER analyses~\cite{Vinciguerra:2023qxq}, the \texttt{ST+PDT} configuration is statistically preferred when all data—including gravitational wave, radio, X-ray, and theoretical constraints—are considered within a unified EOS inference framework.

\subsection{Comparison with other works}
Comparative analyses between \texttt{ST+PDT} and \texttt{PDT-U}  have been reported by other authors in the context of different underlying EOS models \cite{Rutherford:2024srk, Li:2024pts, Li:2025oxi, Passarella:2025zqb}. Among these, \textcite{Rutherford:2024srk} used piecewise polytrope and speed of sound parametrizations to investigate the effect of these two solutions on the properties of NS. They found some differences in key quantities like $R_{1.4}$ and $M_{\text{max}}$, given their implementation of the $\chi$EFT constraints and the choice of EOS models. In comparison, we find much bigger differences in these quantities. This difference in results may be attributed to our improved treatment of the EOS where we used a hybrid approach of nuclear parametrization around saturation density and piecewise polytropes at higher densities, implementation of $\chi$EFT constraints, inclusion of PREX-II, CREX, pQCD, and the different mass distribution coming from 70 radio pulsar masses. In Refs. \cite{Li:2024pts, Li:2025oxi}, the authors used a covariant density functional EOS model and included PSR J1231$-$1411 data in all their results, making a direct comparison with our result difficult. In Ref. \cite{Passarella:2025zqb}, the author also used a relativistic nuclear model and showed a comparison between \texttt{ST+PDT} and \texttt{PDT-U} to infer the saturation properties. But, they did not provide any comparison for the high density EOS or any NS properties. Regarding the effect of  PSR J0614$-$3329, \textcite{Mauviard:2025dmd} found mild softening of the EOS and a shift in their mass-radius posterior to lower radii values, similar to our results. However, they also reported a bimodality in their $R_{1.4}$ distribution that we have not seen in our case. In another work, \textcite{Ng:2025wdj} have recently found a stronger softening of the EOS within their semiparametric EOS framework after including the data from PSR J0614$-$3329 and PSR 0437+4715.

\section{Conclusion}
\label{section: conclusion}

We have extended a multi-messenger Bayesian framework for neutron star EOS inference by incorporating the recent NICER mass–radius measurement of PSR J0614$-$3329 and evaluating alternative pulse profile models for PSR J0030+0451. Our analysis shows that the inferred EOS and neutron star properties—such as radii, tidal deformabilities, and the maximum mass—can depend sensitively on the geometric modeling of surface hotspots in NICER data. We find strong Bayesian evidence favoring the simpler two-spot model (\texttt{ST+PDT}) over the more flexible shape-unconstrained model (\texttt{PDT-U}) for PSR J0030+0451. This result illustrates that multi-messenger EOS inference can act as a diagnostic tool for assessing internal consistency across different observational modeling strategies. We also included the recent mass–radius measurement of PSR J0614$-$3329 in our analysis. Its inclusion leads to a modest tightening of the overall EOS posterior and shifts the inferred radii toward the softer side by approximately 100 meters. 

As the quality and quantity of neutron star observations continue to improve, joint inference frameworks such as the one used here will become increasingly important-not only for constraining the EOS but also for ensuring consistency among data sources and modeling assumptions. This work highlights the value of combining astrophysical, theoretical, and experimental inputs into a single coherent analysis.

\section*{Acknowledgements}
  BB  acknowledges the support from the Alexander von Humboldt Foundation through a Humboldt Research Fellowship for Postdoctoral Researchers. Calculations were performed on the facilities at the SUNRISE HPC facility supported by the Technical Division at the Department of Physics, Stockholm University. This project has received funding from the European Union’s Horizon 2020 research and innovation programme under the Marie Skłodowska-Curie grant agreement No. 101034371. PC acknowledges the support from the European Union's HORIZON MSCA-2022-PF-01-01 Programme under Grant Agreement No. 101109652, project ProMatEx-NS.

\bibliography{mybiblio}
\end{document}